\documentclass[11pt]{article}
\usepackage{graphicx,floatflt,amssymb,epsfig}
\usepackage{rotating}
\def\gtap{\raisebox{-.4ex}{\rlap{$\sim$}} \raisebox{.4ex}{$>$}}

\def\Gn{G_{\vec{n}}}
\def\mnsq{m_{\vec{n}}^2}
\def\mn{m_{\vec{n}}}

\begin{document}
\begin{flushright}
\texttt{hep-ph/0512050}\\
SINP/TNP/05-28\\
TIFR/TH/05-46\\
\end{flushright}

\vskip 20pt

\begin{center}
{\Large \bf Studying the effects of minimal length in large extra
dimensional models in the jet $\mathbf +$ missing energy channels at
hadron colliders} \\
\vspace*{1cm} \renewcommand{\thefootnote}{\fnsymbol{footnote}} {\large
{\sf Gautam Bhattacharyya ${}^1$}, {\sf
Kumar Rao ${}^2$}, {\sf K. Sridhar ${}^2$} } \\
\vspace{10pt}

{\small  1. Saha  Institute  of Nuclear  Physics,  1/AF Bidhan  Nagar,
        Kolkata 700064, India \\ 2. Department of Theoretical Physics,
        Tata  Institute of Fundamental  Research, \\Homi  Bhabha Road,
        Mumbai 400005, India }

\normalsize
\end{center}

\begin{abstract}

\noindent 
Theories of quantum gravity suggest the existence of a minimal length scale.
We study the consequences of a particular implementation of the idea of a
minimal length scale in the model of large extra dimensions, the ADD model. To
do this we have looked at real graviton production in association with a jet
at hadron colliders. In the minimal length scenario, the bounds on the
effective string scale are significantly less stringent than those derived in
the conventional ADD model, both at the upgraded Tevatron and at the Large
Hadron Collider.

\vskip 5pt \noindent
\texttt{PACS Nos:~ 11.25.Wx, 13.85.Qk} \\
\texttt{Key Words:~~Extra 
dimension, Minimal length, Hadron Collider}
\end{abstract}

\renewcommand{\thesection}{\Roman{section}}
\setcounter{footnote}{0}
\renewcommand{\thefootnote}{\arabic{footnote}}

\section{Introduction}
There are reasons to believe that at the Planck scale, the scale at which
gravity becomes a quantum phenomenon, the very structure of space-time
may change. That this may happen is suggested even by General Relativity.
A quantum mechanical particle of momentum $p$ in the presence of a classical
gravitational field (the latter described by Einstein's equations) causes
the metric $g$ to fluctuate. This induces an additional 
uncertainty in position,  
given by $l_p^2 \Delta p$, where $l_p$ is the Planck
length. Thus the usual quantum mechanical uncertainty relation gets
modified to
\begin{equation}
\Delta x ~\gtap~ {1 \over \Delta p} +  l_p^2~ \Delta p.
\end{equation}
At high energies, the second term can become significant and lead to important
deviations from the usual quantum mechanics. For example, in the usual quantum
mechanics, $\Delta x$ is large at low values of momenta but can become small
at high momenta which can provide higher resolution. With the modified
uncertainty relation even at high momenta $\Delta x$ is limited in resolution
because of strong curvature effects. In other words, independent of momentum,
$\Delta x$ is always larger than a minimal length scale $l_p$. The appearance
of the minimal length in the classical theory of gravity should tell us that
it is no surprise to expect that such a conclusion becomes even more
inevitable in a quantum theory of gravity. Indeed, a whole range of quantum
gravity models predict the existence of a minimal length \cite{garay, ng}.
However, because the Planck length is so minuscule it is of consequence only
to the physics of the early universe, if at all \cite{hassan}. There is
a twist in the plot, however, and that is due to the development of
brane-world physics. This new paradigm in physics has far-reaching
implications but of our immediate interest is the fact that there are models
of brane-world which allow a realisation of quantum gravity effects at very
low energy scales -- as low as a TeV.  In this paper, we will deal with the
simplest of such realizations which is a model due to Arkani-Hamed, Dimopoulos
and Dvali.

In the model of Arkani-Hamed, Dimopoulos and Dvali, the so-called ADD model
\cite{lxd}, extra space dimensions are compactified to have large volume moduli
and thereby it becomes possible to lower the scale of quantum gravity from the
Planck scale to the TeV scale.  In the ADD model, the gauge interactions are
confined to a 3-brane while gravity propagates in all the available
dimensions. The higher dimensional Planck scale $M_S$ in the ADD model is
related to the usual Planck scale by $M_{\rm{Pl}}^{2}= R^{d} M_{S}^{d+2}$,
where $R$ is the radius of compactification and $d$ is the number of extra
space dimensions. For appropriate choices of $R$ and $d$, the fundamental
(higher dimensional) Planck or string scale can be brought down to $M_S \sim$
1 TeV. This scale can be probed at present and future colliders. Various
signals of graviton production and virtual graviton exchange have been studied
\cite{hlz, grw} and reviewed \cite{perez,sridhar} in recent years. In the
context of ADD model with $M_S \sim 1$ TeV, the minimal length hypothesis is
phenomenologically interesting if we take it to be around an inverse TeV, viz.
$l_{p} \sim 1/M_S$. For a review, see Ref.~\cite{hossenfelder_review}.

Different applications of the minimal length scenario (MLS) have been
discussed in \cite{hossen1,restml,bmks}.  In an earlier analysis \cite{bmks},
we had studied how the MLS hypothesis influences the dilepton production
process in hadron colliders.  In the ADD model, for such processes involving
virtual gravitons, one has to sum over an infinite tower of graviton
Kaluza-Klein (KK) states at the amplitude level. The result is divergent and
to cure the divergence an {\it ad-hoc} cutoff of the order of $M_{S}$ is used.
However, in the MLS scenario, the minimal length acts as an ultraviolet
regulator and allows one to sum over the entire KK graviton tower by smoothly
cutting off the contribution of higher energy KK states rendering the
amplitude finite. Another important modification comes from the rescaling of
momentum measure leading to an alteration in the phase space integration.
These, as we observed in \cite{bmks}, lead to a significant deviation of the
bound on $M_S$ from the one obtained in the conventional ADD picture without
the MLS hypothesis.

In the present analysis, we consider processes involving emission of real
gravitons in hadron colliders, namely, the upgraded Tevatron and the future
Large Hadron Collider (LHC). The partonic sub-processes are $P_1 P_2 \to P_3
\Gn$, where $P_i$ are the appropriate quarks/gluons. Here $\Gn$ means a real
graviton with KK index $\vec{n} = (n_1, n_2, \cdot \cdot, n_d)$, where $d$ is
the number of extra dimensions. For a given process, one must sum over such
external modes at the cross section level. This KK summation gets modified in
the MLS scenario. The phase space integration also gets modified in the MLS
scenario. Again, as we shall see towards the end, the bounds that one obtains
on $M_S$ in the conventional ADD model from the jet plus missing energy mode
in the hadron colliders are considerably weakened if we implement the MLS
hypothesis.

\section{The MLS scenario and the jet $\mathbf +$ missing energy channel}
Specifically, by MLS scenario we mean the ADD model with 
a minimal length $l_p$ of the order of TeV${}^{-1}$ incorporated in it.
Since the uncertainty in position measurement now cannot be smaller
than $l_p$, now the standard commutation relation between position and
momentum has to be modified \cite{kempf}.  In the MLS scheme, the Compton
wavelength ($\lambda = 2\pi/k$) of a particle cannot be arbitrarily
small.  We suppose that even though the wave vector $k$ is bounded
from above, the momentum $p$ can be as large as possible. In the same
way, the frequency $\omega$ is restricted from above while the energy
$E$ can go up arbitrarily. This means that the standard relations
$p=\hbar k$ and $E=\hbar \omega$ need to be modified. This can be
realised by introducing the Unruh ansatz \cite{unruh}
\begin{eqnarray}
\label{dispersion}
l_{p}k(p)&=& \tanh^{1/\gamma}
\left[\left(\frac{p}{M_{S}}\right)^{\gamma}\right],  \nonumber \\ 
l_{p} \omega(E)&=&\tanh^{1/\gamma}
\left[\left(\frac{E}{M_{S}}\right)^{\gamma}\right],
\end{eqnarray}
where $\gamma$ is a positive constant. Eq.~(\ref{dispersion})
captures the essence of a minimal length scale: as $p$ (or $E$) becomes very
large, $k$ (or $\omega$) approaches the upper bound $1/l_{p}$ and hence one
cannot probe arbitrarily small length and time scales.  The generalized
position-momentum and energy-time uncertainty principle can now be written in
a Lorentz covariant form as
\begin{equation}
\label{ur}
\left[x^{\nu},  p_{\mu}\right]=i  \,\,\frac{\partial p_{\mu}}{\partial
k_{\nu}}. 
\end{equation}
As explained in \cite{kempf,hossen1}, the effect of Eq.~(\ref{dispersion}) can
conveniently be accounted for by a simple redefinition of the momentum
measure, which in one dimension is given by
\begin{equation}
dp \longrightarrow dp \,\,\frac{\partial k}{\partial p}. 
\end{equation}
The Lorentz invariant 4-momentum measure is modified as
\begin{equation}
\label{scale}
d^{4}p    \to     d^{4}p    \,\,\,\textrm{det}    \left(\frac{\partial
k_{\mu}}{\partial  p_\nu}\right)=d^{4}p  \prod  _{\nu}  \frac{\partial
k_{\nu}}{\partial p_\nu}, 
\end{equation}
where the Jacobian matrix $\frac{\partial k_{\nu}}{\partial p_{\nu}}$
can be kept diagonal.

In this paper we study how processes involving real graviton emissions in
hadron colliders are going to be influenced by the MLS hypothesis. The emitted
gravitons will disappear carrying missing energy, so the events that we look
for are jets plus missing energy. The partonic sub-processes that contribute
to this channel are $q\bar{q} \to g \Gn$, $qg \to q \Gn$, $\bar{q}g \to
\bar{q} \Gn$, and $gg \to g \Gn$. The differential cross section for a
particular sub-process (labelled $i$) for a given KK graviton mode has the
form
\begin{equation}
\label{ds}
\frac{d\sigma_i^{\vec{n}}}{d\cos\theta} = G_N f_i(\mnsq, s,t,u),
\end{equation}
where $m_{\vec{n}}$ is the mass of $\Gn$, and $f_i$ are functions of the
Mandelstam variables $s,t,u$ whose explicit expressions are given in
Ref.~\cite{grw,peskin}. These cross sections have to be convoluted with the
density of KK states given by \cite{hlz}
\begin{equation}
\rho(\mn) = \frac{R^d \mn^{d-2}}{(4\pi)^{d/2} \Gamma(d/2)}.
\end{equation} 
Noting that one can express $G_N = (4\pi)^{d/2} \Gamma(d/2)/(2 M_S^{d+2}
R^d)$, the differential cross section summed over the kinematically allowed KK
graviton tower is given by
\begin{equation} 
\label{dscon}
\frac{d\sigma^{\rm ADD}_i}{d\cos\theta} = \frac{1}{2 M_S^4} \int d\mnsq
 \left(\frac{\mnsq}{M_S^2}\right)^{\frac{d-2}{2}} f_i(\mnsq, s,t,u) ,
\end{equation}
where the lower limit of integration is zero and the upper limit corresponds
to the maximum allowed graviton energy as $\sqrt{s}/2$.  An important thing
that has to be noticed here is that while the usual 4-dimensional graviton
emission cross section goes like $1/M_{\rm Pl}^4$, the effect of KK summation
jacks up the contribution which goes effectively as $1/M_S^4$. This is the ADD
effect. Now, how does the MLS hypothesis modify the KK summation?  Now for
analytic simplicity we take $\gamma=1$ and set $l_{p} M_{S} = \hbar = 1$. The
departure from these assumptions will be discussed later. The Unruh relations
(\ref{dispersion}) will modulate the integral in (\ref{dscon}) by a factor
$\partial \omega/\partial E$. The MLS-influenced differential cross section is
given by
\begin{equation} 
\label{dsmls}
\frac{d\sigma^{\rm MLS}_i}{d\cos\theta} = \frac{1}{2 M_S^4} \int d\mnsq
~\left(\frac{\mnsq}{M_S^2}\right)^{\frac{d-2}{2}} 
~{\rm sech}^2\left(\frac{\mn}{M_S}\right)
f_i(\mnsq, s,t,u) .
\end{equation}
For large argument, the sech-square function falls exponentially, thus the
contributions from high mass KK states will be smoothly cut off. 

Another important modification due to the MLS is the change in the cross
section due to the rescaling of the momentum measure given in
Eq.~(\ref{scale}). It is straightforward to check, as derived in
\cite{hossen1}, that the phase space integration in the total cross section
picks up the following modification factor:
\begin{equation} 
\label{mod}
d\sigma^{\rm MLS} = d\sigma^{\rm ADD}~\prod_n \frac{E_n}{\omega_n} ~
\prod_\nu  \left. \frac{\partial k_\nu}{\partial  p_\nu}\right|_{p_i =
p_f},
\end{equation}
where $n$ runs over the four initial and final states in a $2 \to 2$ process,
and $p_i$ and $p_f$ are the total initial and final four momenta in a $2 \to
2$ process.  We work out this modification factor for the process we are
studying viz. $P_1 P_2 \rightarrow P_3 \Gn$. Note, while $P_i$ are massless,
$\Gn$'s are massive.  Using Eqs.~(\ref{dispersion}) and (\ref{scale}), setting
$\gamma = 1$ and $l_p M_S = 1$, we can easily show that
\begin{eqnarray} 
\label{ebyomega}
\prod_n \frac{E_n}{\omega_n} & = & \frac{s p_T m_T}{4M_S^4} 
~\textrm{coth}\left\{\frac{p_{T}\cosh(y_1)}{M_S}\right\}
~\textrm{coth}\left\{\frac{m_{T}\cosh(y_2)}{M_S}\right\}
\prod_{i=1,2}
x_i ~\cosh(y_i) ~\textrm{coth}\left(\frac{x_i\sqrt{s}}{2M_S}\right), 
\nonumber \\ 
\prod_\nu
\frac{\partial k_\nu}{\partial p_\nu} & = & {\rm   sech}^2
\left\{\frac{(x_1+x_2)\sqrt{s}}{2M_S} \right\} {\rm sech}^2
\left\{\frac{(x_1-x_2)\sqrt{s}}{2M_S} \right\}.
\end{eqnarray} 
Above, $x_1$ and $x_2$ are the momentum fractions of the hadrons carried by
the initial state partons, $y_1$ and $y_2$ are the rapidities of the final
state parton and the graviton, $p_T$ is the transverse momentum of the final
state parton, and $m_T = \sqrt{p_T^2 + \mnsq}$ is the transverse mass of the
graviton $\Gn$.  It is instructive to check that in the decoupling limit $M_S
\gg \sqrt{s}$, the phase space correction factor goes to unity.

The sech-square modulation in Eq.~(\ref{dsmls}) together with the phase space
modifications in Eqs.~(\ref{mod}) and (\ref{ebyomega}) constitute the complete
correction. If we keep $\gamma$ and $\delta = 1/(l_p M_S)$ as free parameters
the analytic expressions of the modifications in Eqs.~(\ref{dsmls}) and
(\ref{ebyomega}) will look more complicated, and we do not display
them. Instead, we have numerically demonstrated the sensitivity of cross
section due to variations of $\gamma$ and $\delta$ in the next section.

\section{Results}
We have numerically computed the cross section for the production of the
graviton KK tower in association with a jet at the Run II of Tevatron
($\sqrt{s}=$ 1.96 TeV) and at the LHC ($\sqrt{s}=$ 14 TeV) in both the ADD
model and in the MLS scenario.  There is an irreducible SM background to the
signal coming from $Z+$jet production followed by the invisible decay of the
$Z$ into neutrinos. We have computed the SM background and the signal in the
ADD and the MLS cases using the MRST 2001 LO parton distributions \cite{mrst}
(the MLS modification of the parton densities is numerically insignificant).
Since the signal peaks at larger missing-$E_T$ as compared to the background
the purity of the sample can be improved by employing a strong cut on the
missing-$E_T$. We require the missing-$E_T$ to be greater than 80 GeV for the
Tevatron Run II analysis. The cut on the jet pseudorapidity is $\vert \eta_j
\vert <$ 2.4. Our results are shown for the case $d=3$.  We display the
results of this computation in Figure 1 by plotting the cross section in the
ADD and the MLS scenarios as a function of $M_S$.  Experimentally, the $Z+$jet
cross section is not known very precisely. To have a rough estimate, we
multiplied the SM cross section by the appropriate luminosity to obtain the
number of events, and twice the square-root of those events (back to pb unit)
correspond to the 2-$\sigma$ spread around the SM value. Thus, no detector
effect has been taken into consideration.  In other words, assuming purely
statistical errors on the SM cross section, we have plotted the 2-$\sigma$
allowed bands in the figure, which may be considered as a rough estimate of
the allowed zone.  As expected, the MLS cross section has a steeper fall-off
with $M_S$.  Consequently, the lower bound on $M_S$ in the MLS scenario is
diluted by about 60 GeV compared to what one obtains in the conventional ADD
model at the $95\%$ confidence level.

For the LHC analysis, we have used the much stronger cut of 200 GeV on the
missing $E_T$ and a jet pseudorapidity cut of $\vert \eta_j \vert <$ 3.5.  The
cross section for the signal in the two cases and the SM-allowed band are
shown in Figure 2.  The effects of the suppression of cross section due to the
minimal length is stronger at the LHC and hence the dilution of the bound (by
a few hundred GeV) is greater than at the Tevatron.

Figures 1 and 2 have been drawn setting $\gamma = 1$ and $l_p M_S = 1$. In
Figure 3, we have displayed the variation of the total cross section, for
Tevatron RUN II, with $\gamma$ for different choices of $\delta = 1/(l_p
M_S)$. The value of $M_s$ has been fixed at 1.5 TeV for this figure. In Figure
4, we have presented the analogous variation for the LHC energy for $M_S = 7$
TeV.  We recall that the dependence of the cross section on $\gamma$ and
$\delta$ arise from the use of Eq.~(\ref{dispersion}) while calculating the
MLS modulation factors.

\section{Discussions and Conclusions}
The existence of a minimal length is a generic prediction of quantum gravity
theories. The model of large extra dimensions, the ADD model, is an example
where quantum gravity effects become manifest at energy scales as low as a TeV
making it accessible to collider experiments. We incorporate the idea of a
minimal length in the ADD model, which we now call the MLS scenario, and study
what impact the existence of a minimal length has on the collider effects of
the conventional ADD model.  In particular, we have considered graviton
production in association with jets at hadron colliders to address this
particular issue.  By analysing the jet$+$missing energy channel at the
upgraded Tevatron and the Large Hadron collider, we demonstrate that the ADD
model bounds get diluted when the minimum length hypothesis is invoked.

Finally, we make a few comments on the possible implications of minimal length
on the astrophysical constraints on the ADD model.  First we note that
gravitons produced in supernovae core by nucleon nucleon bremsstrahlung
process can carry away missing energy.  But neutrino fluxes from SN1987A
detected by the IMB and Kamiokande experiments have been observed to carry
away most of the energy ($E \ge 2 \times 10^{53}$ ergs) released during core
collapse.  This has been used to obtain tight bounds on $M_S$ of the order of
50 TeV for $d = 2$ and about 4 TeV for $d = 3$ \cite{supernova} in the
conventional ADD model. Since the production process of the gravitons is the
same as the one studied in the present paper, it is important to ask as to
what extent the MLS scenario dilutes the supernova bounds.  Note, the high
nucleon density at the supernova core and the large multiplicity of KK
graviton modes for low values of $d$ is responsible for the enhancement of the
effects, which in turn places strong constraints on $M_S$.  The MLS scenario
would indeed suppress the graviton production cross section and tend to dilute
the bounds.  But the numerical impact of this dilution would not be large
compared to the other uncertainties, e.g. strong core temperature dependence,
inherited from pure supernova dynamics.

A string theory calculation of this cross section has been done in
\cite{string}, where the amplitude is multiplied by a form factor. It is worth
comparing the string theory result with the Unruh ansatz that we have used. In
the string theory case the form factor goes like ${\rm exp} (-m_n^2/M_S^2)$ in
the asymptotic limit. In our case, the modulation depends on the choice of
$\gamma$.  For $\gamma = 1$, the high mass KK states are cut off by ${\rm exp}
(-m_n/M_S)$ since the argument of sech-square in Eq.~(\ref{dsmls}) has a
linear dependence on graviton mass. For $\gamma = 2$, the dependence on the
graviton mass is quadratic, and we obtain a string-like modulation.  We note
at this point that the string calculation as done in \cite{string} is based on
a toy scenario with an embedding of QED in string theory, and the calculation
of jet plus missing energy relies on a rough extrapolation from this toy
scenario. Because of this model dependence, a more detailed comparison of the
results in \cite{string} and our analysis is not possible.

\section*{Acknowledgements}
GB and KS acknowledge support by the Indo-French Centre for the Promotion of
Advanced Research, New Delhi (IFCPAR Project No: 2904-2), and thank CERN PH/TH
division for hospitality during the work. GB also thanks LPT-Orsay and
SPhT-Saclay for hospitality during the final stage of the work. GB's research
has also been supported, in part, by the DST, India, Project No:
SP/S2/K-10/2001.

\newpage

\begin{figure}    
 \centering    
%\vspace{-1cm}  
%\begin{tabular}{cc}      
\includegraphics[scale=0.5, angle=-90]{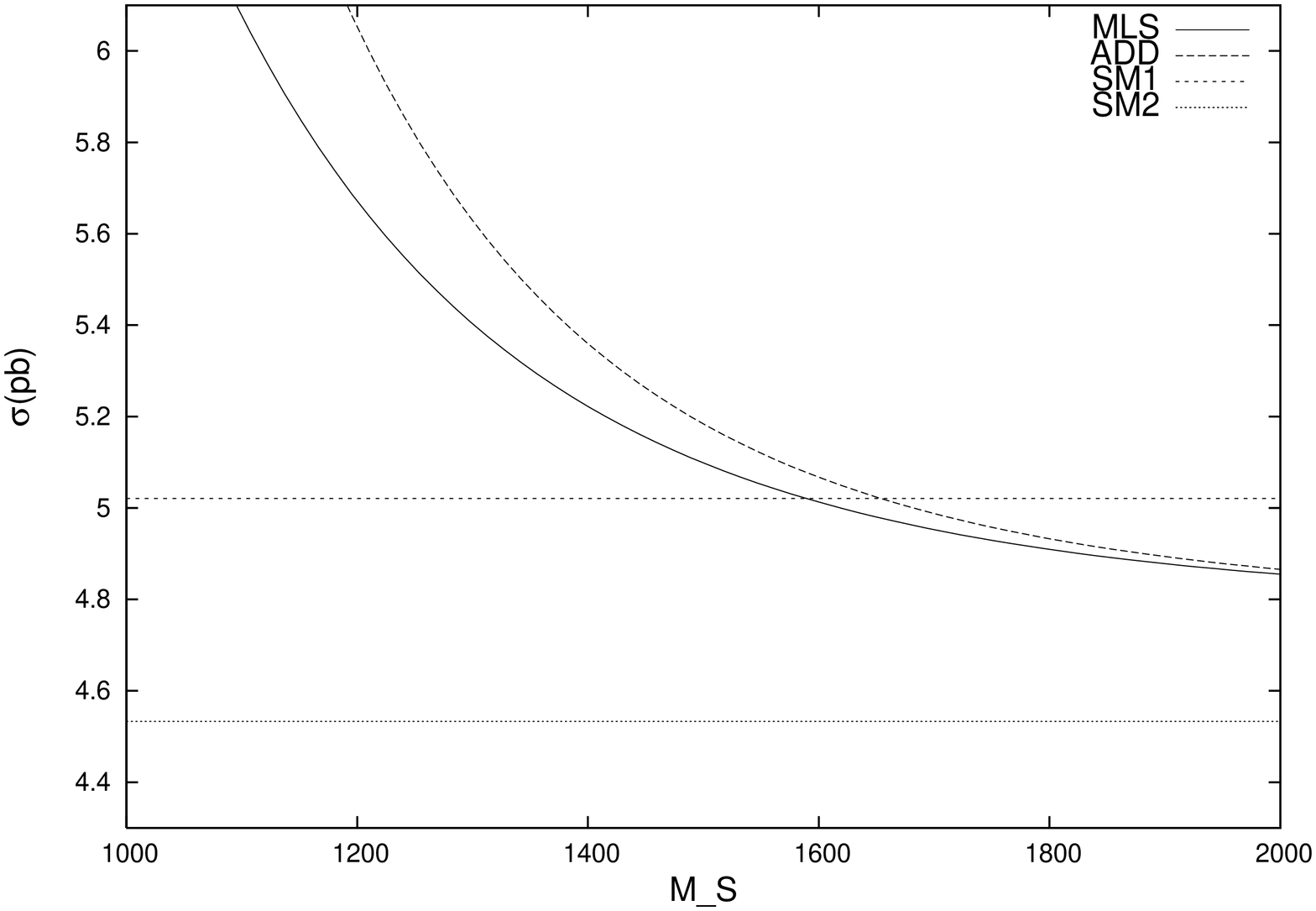}     
%\\     
%\end{tabular}     
\caption{The jet$+$missing energy cross section as a function of $M_S$ for the
  Tevatron RUN II ($\sqrt{s}=$ 1.96 TeV).  The two curves are for the
  conventional ADD model and the MLS scenario, assuming the number of extra
  dimensions $d = 3$.  Also shown are the 95\% C.L. upper and lower bounds
  (`SM1' and `SM2', respectively) on the SM cross section (assuming only
  statistical errors).}  \protect\label{fig1}
\end{figure}

\begin{figure}   
 \centering    
%\hspace{-1cm}  
%\begin{tabular}{cc}      
\includegraphics[scale=0.5, angle=-90]{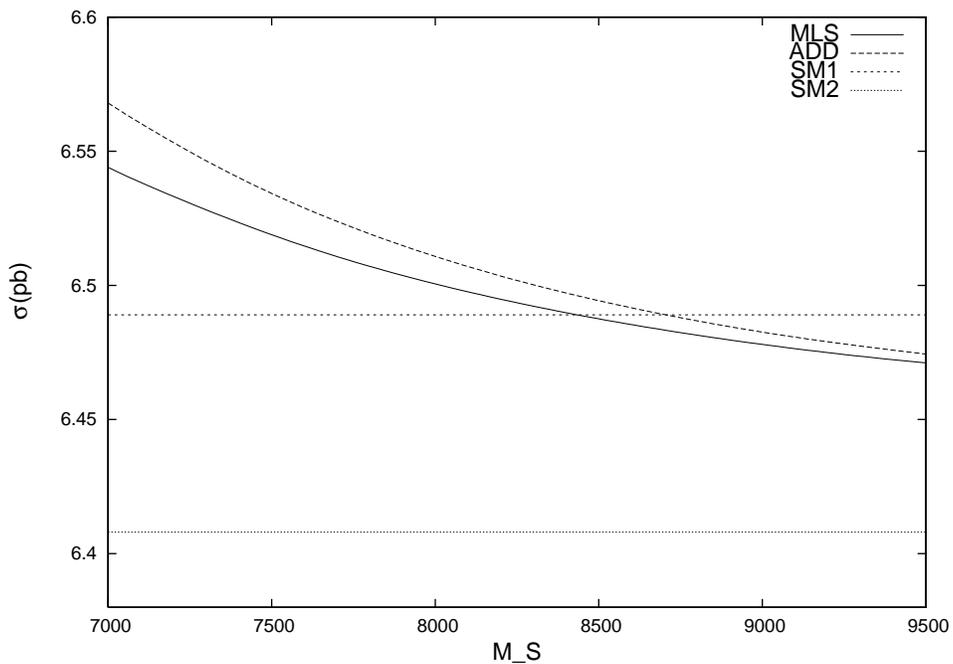}     
%\\     
%\end{tabular}     
\caption{Same as in Figure 1, but for the LHC ($\sqrt{s}=$ 14 TeV).}
\protect\label{fig2}
\end{figure}

\begin{figure}
 \centering    
%\hspace{-1cm}  
%\begin{tabular}{cc}      
\includegraphics[scale=0.5, angle=-90]{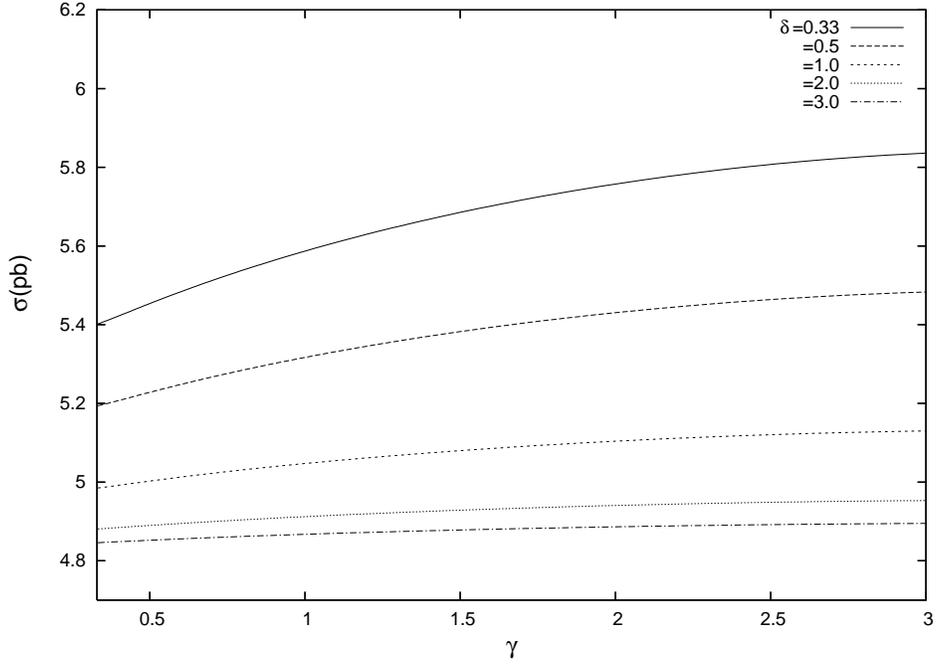}     
%\\     
%\end{tabular}     
\caption{Variation of the MLS cross section with $\gamma$ for different values
  of $\delta = 1/(l_p M_S)$ in the context of Tevatron RUN II ($\sqrt{s}=$
  1.96 TeV). The plots correspond to the fixed $M_S = 1.5$ TeV.}
  \protect\label{fig3}
\end{figure}

\begin{figure}   
 \centering    
%\hspace{-1cm}  
%\begin{tabular}{cc}      
\includegraphics[scale=0.5, angle=-90]{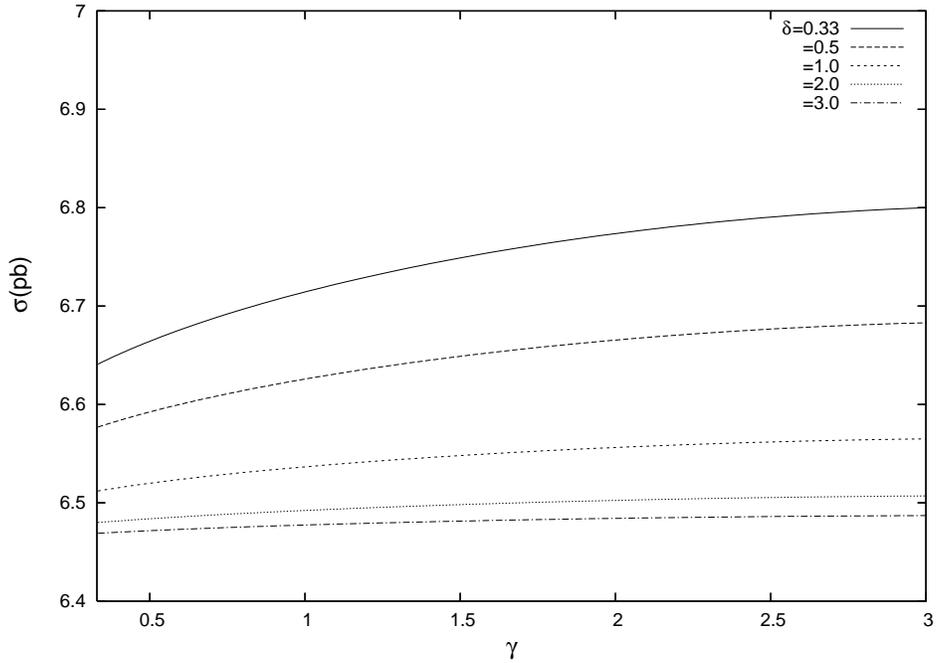}     
%\\     
%\end{tabular}     
\caption{Same as in Figure 3, but for the LHC ($\sqrt{s}=$ 14 TeV), with fixed
  $M_S = 7$ TeV.}  \protect\label{fig4}
\end{figure}

\end{document}